\documentclass[fleqn,usenatbib]{mnras}

\usepackage{newtxtext,newtxmath}

\usepackage[T1]{fontenc}

\usepackage{soul}

\DeclareRobustCommand{\VAN}[3]{#2}
\let\VANthebibliography\thebibliography
\def\thebibliography{\DeclareRobustCommand{\VAN}[3]{##3}\VANthebibliography}

\usepackage{graphicx}	
\usepackage{lipsum}
\usepackage{caption}
\usepackage{subcaption}

\title[Variability as low-mass AGN]{Photometric variability in star-forming galaxies as evidence for low-mass AGN and a precursor to quenching}

\author[C. Cleland \& S. L. McGee]{
Cressida Cleland,\thanks{E-mail: cressidac@star.sr.bham.ac.uk}
Sean L. McGee
\\
School of Physics and Astronomy, University of Birmingham, Edgbaston, Birmingham, B15 2TT, UK\\
}

\date{Accepted XXX. Received YYY; in original form ZZZ}

\pubyear{2021}

\begin{document}
\label{firstpage}
\pagerange{\pageref{firstpage}--\pageref{lastpage}}
\maketitle

\begin{abstract}
We measure the optical variability in $\sim16500$ low-redshift ($z \sim 0.1$) galaxies to map the relations between AGN activity and galaxy stellar mass, specific star-formation rate, half-light radius and bulge-to-total ratio. To do this, we use a reduced $\chi^2$ variability measure on $>$ 10 epoch lightcurves from the Zwicky Transient Facility and combine with spectroscopic data and derived galaxy parameters from the Sloan Digital Sky Survey. We find that below stellar mass of 10$^{11}$ M$_\odot$,  galaxies classed as star-forming via the BPT diagram have higher mean variablities than AGN or composite galaxies. Revealingly, the highest mean variabilities occur in star-forming galaxies in narrow range of specific star-formation -11 $<$ log$\left(\text{sSFR/yr}^{-1}\right) <$ -10. In very actively star-forming galaxies $\left(\log\left(\text{sSFR/yr}^{-1}\right) > -10\right)$, the reduced variability implies a lack of instantaneous correlation with star-formation rate. Our results may indicate that a high level of variability, and thus black hole growth, acts as a precursor for reduced star-formation, bulge growth, and revealed AGN-like emission lines.  These results add to the mounting evidence that optical variability can act as a viable tracer for low-mass AGNs and that such AGNs can strongly affect their host galaxy.
\end{abstract}

\begin{keywords}
galaxies:evolution -- galaxies:star-formation -- galaxies:active 
\end{keywords}



\section{Introduction}

In the past two decades, the study of active galactic nuclei (AGN) and their properties has become widespread, in part thanks to recent advances in both multi-wavelength astronomy and long-term time-domain surveys. One of the celebrated facets of AGN behaviour is that the central black hole is well correlated with the growth of the bulge, and thus, the growth of the galaxy stellar mass \citep{Magorrian1998, Mcconnell2013, Kormendy2013, Reines2015}. This tight correlation implies a self-regulation which links galaxy evolution to black hole growth that has been shown to be important in theoretical models of galaxy formation \citep{dimatteo05, Croton2006, Bower2006}. This raises the question - can we see observational evidence that AGN activity and the resulting black hole growth has a direct effect on galaxy properties such as its star-formation rate or morphology?

Theoretically, it is understood that AGN feedback has an effect on the star-formation rate of a galaxy, by either heating up cold gas and thus preventing it from condensing to form stars, or by energetically expelling gas from star-forming regions \citep[see][for a review]{Fabian2012}. However, these processes are still not fully understood and this mode of quenching is heavily confounded with other processes such as environmental quenching \citep[e.g.][]{Wetzel2013,Cleland2021} and/or mass quenching \citep{peng10, Peng2015}. Therefore, it is useful to investigate AGN activity in galaxies across the full stellar mass range in a wide variety of environments and stages of evolution.

Unfortunately, it is difficult to find a single observational measure which gives a uniform accounting of the presence of an AGN. Generally, commonly used indicators cannot be uniformly applied to all galaxies with equal effectiveness. For instance, the often-used practise of measuring emission line ratios to distinguish AGN emission from star-formation is often not effective in strongly star-forming galaxies because the star-formation swamps the AGN emission \citep{baldwin81, Cann2019, Agostino2021}. One promising route to a uniform indicator is through examining the photometric variability. Variability in the luminosity of galaxies can act as an indicator for varying activity levels in the AGN. Higher levels of activity over the observing period may suggest stronger black hole growth and thus stronger evolutionary links. In the short term, i.e. over timescales of days to years, variability has been detected in lightcurves of AGN in a variety of wavelengths \citep{cartier15,caplar17,smith18}. Variability over longer timescales has been uncovered through the use of models and simulations \citep{novak11,blandhawthorn13, Sartori2018}. Optical variability has also been shown to be a reliable method of uncovering low-mass AGN otherwise obscured by star-formation \citep{baldassare18,baldassare20, ward2021, Burke2021}. 

In this paper we attempt to extend the application of optical variability across a full sample of galaxies at low redshift drawn from the Sloan Digital Sky Survey. To quantify the optical variability, we use the reduced $\chi^2$ of the magnitudes of galaxy lightcurves drawn from the Zwicky Transit Facility. This is found to be a robust method, both from our own formalism, and from the work of \citet{zibecchi20}. This method is agnostic of galaxy type and can be applied to the entire sample, including galaxies that are deemed non-AGN by other methods such as a BPT diagram \citep{baldwin81}. In $\S$\ref{sec:data} we describe the dataset, in $\S$\ref{sec:analysis} we explain how variability is calculated and discuss some initial results, in $\S$\ref{sec:discu} we interpret and discuss the results, and in $\S$\ref{sec:conc} we summarize. Throughout we adopt the cosmology used in \citet{omand15}, from where we derive our sample \citep[see also][]{bennett13}, with $h=0.693$, $\Omega_M=0.286$, and $\Omega_\Lambda =0.714$ and a \citet{kroupa01} IMF.

\begin{figure*}
     \centering
     \includegraphics[width=\textwidth]{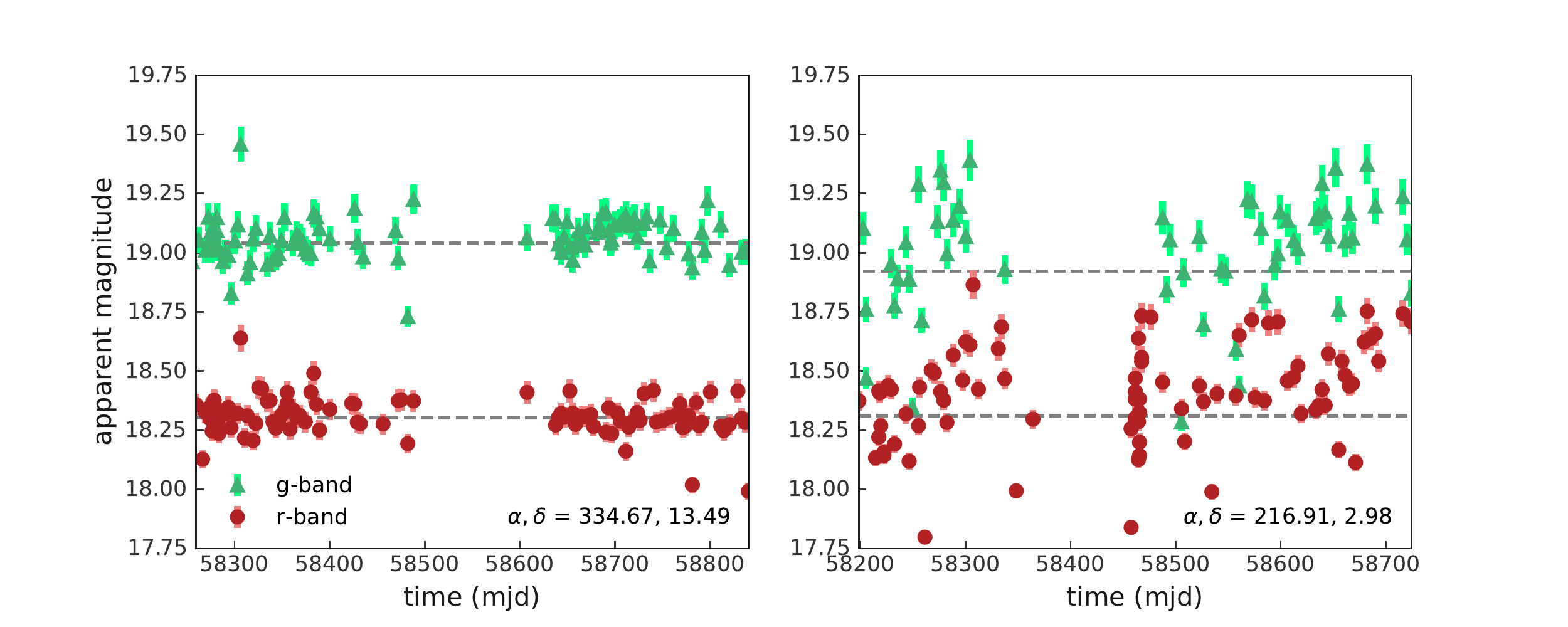}
     \caption{Two examples of galaxy lightcurves, with variability $=0.46$ on the left and variability $=1.53$ on the right as measured using Equation \ref{eqn:var}. {$r$-band magnitudes are plotted as red circles, with $g$-band magnitudes plotted as green triangles, for illustrative purposes. Here, the dashed line represents the constant line of best fit through the photometric datapoints.} These galaxies were chosen to have similar mean $r$ magnitudes, similar number of photometric observations, and similar observation time periods.}
     \label{fig:lightcurve}
\end{figure*}

\section{Data} \label{sec:data}
Our sample is based on galaxies with spectroscopic measurements from the Sloan Digital Sky Survey (SDSS) Data Release 7 with $0.01<z \leq 0.2$ \citep{abazajian09}. We use a sample of galaxies constructed by \citet{omand15}\footnote{\url{http://cdsarc.u-strasbg.fr/viz-bin/cat/J/MNRAS/440/843}}, which contains several value-added catalogues to use in addition, including the stellar mass ($M_*$) and star-formation rate (SFR) which come from the MPA-JHU value-added catalogue \citep{brinchmann04}\footnote{\url{https://wwwmpa.mpa-garching.mpg.de/SDSS/DR7/}}, and effective radii and bulge-to-total estimations from \citet{simard11}\footnote{\url{https://cdsarc.cds.unistra.fr/viz-bin/cat/J/ApJS/196/11}}. The stellar masses are calculated by comparing photometry fits to various stellar population models, and the star-formation rates are inferred from emission lines in the spectra. Using the Zwicky Transient Facility (ZTF; {DR7}) catalogue search\footnote{\url{https://irsa.ipac.caltech.edu/cgi-bin/Gator/nph-scan?projshort=ZTF}}, we match these coordinates to photometric sources observed by ZTF \citep{bellm14,masci19}. We search for matching sources in a radius of 1 arcsecond; this radius was chosen to maximise matches while also avoiding contamination from non-matches. The resulting table provides object IDs that can be used to retrieve photometric lightcurves from ZTF. The photometric lightcurves were measured with psf-matching source extraction at each individual epoch and the range of magnitudes we cover are well detected in each frame. We restrict observation magnitudes to the $r$-band {as it is the same band as the spectroscopic selection. We have aimed to make as homogeneous a galaxy selection as possible, and examining the $r$-band avoids the need for cross-band corrections}.

In total, we retrieve 51068 non-empty lightcurves matched with the \citet{omand15} sample. In Figure  \ref{fig:lightcurve}, we show two example lightcurves {in both the $g$- and $r$-band, although our variability measures use only the $r$-band.} These were selected to highlight the difference between high and low variability (0.46, left; 1.53, right). The observations were taken over similar time periods (581 days, left; 526 days, right), similar number of photometric observations (191, left; 93, right) and similar mean $r$ magnitudes (and thus similar photometric error bars). The variabilities exhibited here are typical of the ones we study in this paper -- that is, they are not the result of systematic trends in the data, or `one-off' transient events. They seem to result from short-timescale apparent stochastic variability. {By comparing the $g$- and $r$-band variability, we can see that there is clearly some very short timescale variation ($<$ week) which appears in one band but not another. For instance, in the right plot of Figure \ref{fig:lightcurve} there is a large dispersion in the $r$ magnitudes near mjd $\sim$ 58475 which do not appear in the $g$ magnitude. Nonetheless, there is some more coherent variation in both magnitude bands near the beginning and end of the light curves which seems of a physical nature.}

We extract emission line fluxes from \citet{brinchmann04} to construct a BPT diagram \citep{baldwin81} in order to classify galaxies between AGN, star-forming, or composite, based on their emission line ratios.

\section{Analysis and Results} \label{sec:analysis}
\subsection{Calculating variability}
\begin{figure}
    \centering
    \includegraphics[width=\columnwidth]{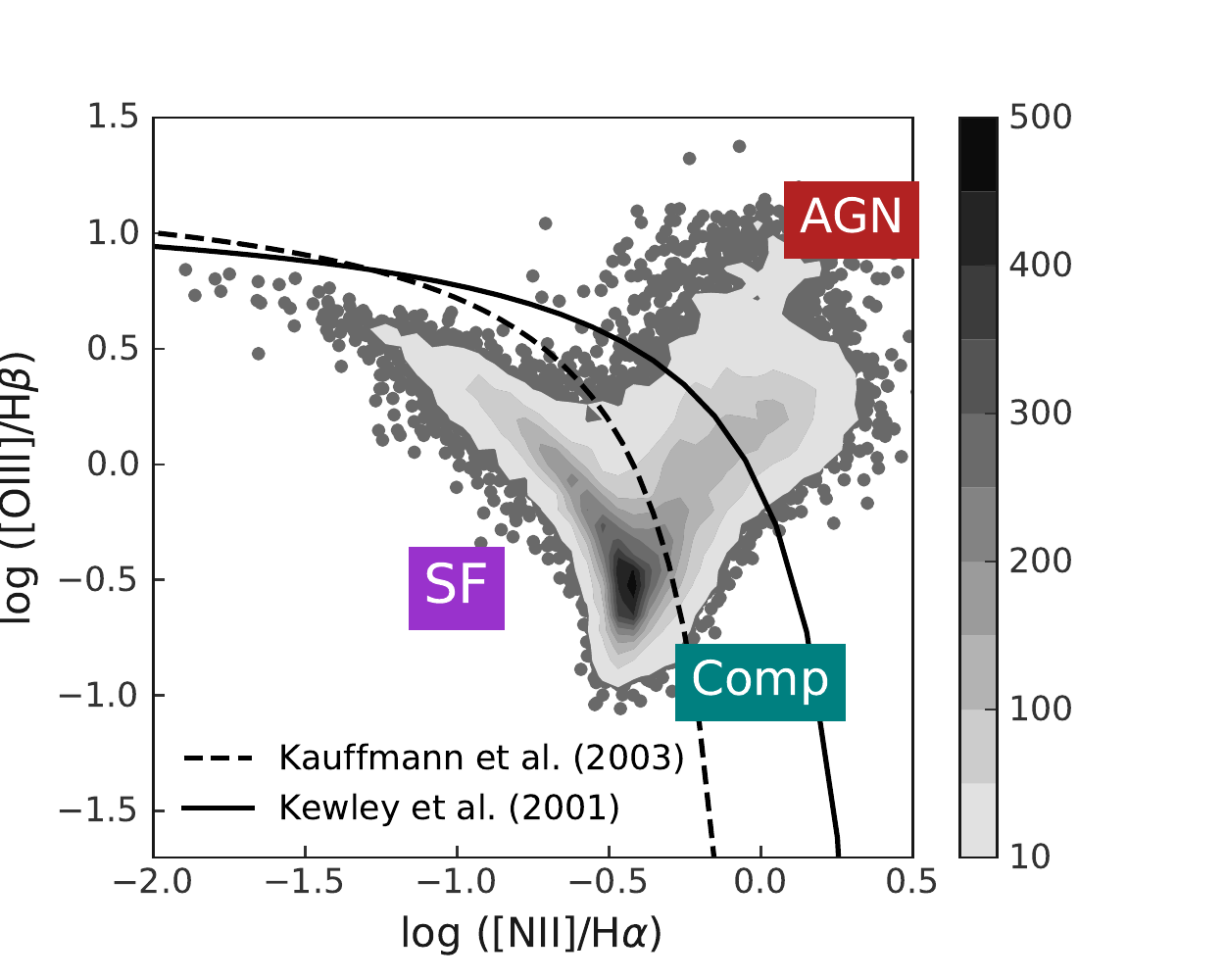}
    \caption{BPT diagram with demarcation lines from \citet{kauffmann03} (dashed) and \citet{kewley06} (solid). The colorbar indicates number of galaxies found at each point in a contour region.}
    \label{fig:bpt}
\end{figure}

To ensure the robustness of the galaxy classes and variabilities, we impose a signal-to-noise cut on the SDSS spectral emission lines and a cut on the number of photometric observations from ZTF. We enact a signal-to-noise ratio cut of $\text{SNR}>3$ for the H$\upalpha$, H$\upbeta$, [NII]$\lambda$6584 and [OIII]$\lambda$5007 emission lines in order to classify between AGN-like emission and emission consistent with star-formation (SF) on a BPT diagram \citep{baldwin81}. The resulting BPT diagram is shown in Figure \ref{fig:bpt}, along with the demarcation lines from the studies of \citet{kauffmann03} and \citet{kewley06}. In the rest of the paper, we will refer to star-forming, composite and AGN, galaxies as labelled in the Figure. For instance, the composite galaxies are those with emission line ratios greater than the Kauffmann line and less than the Kewley line. Note that our classification does not require `star-forming' galaxies to be forming stars at a particular rate, simply that their emission lines appear due to star-formation.

We restrict analysis to lightcurves with $n>10$ photometric epochs from ZTF to ensure robust determination of the variability. However, we have verified that varying the cut-off for $n$ does not quantitatively change the results of this paper. If a galaxy passes the SNR and epoch cuts, the variability is calculated. After SNR and epoch cuts, we are left with 16577 galaxies.

The variability is calculated using the logarithm of the reduced chi-squared $\left(\chi^2_\nu\right)$ of the magnitudes of each epoch, taking the errors on the magnitudes into account, as in Equation \ref{eqn:var}: 
\begin{equation}\label{eqn:var}
\begin{split}
    \chi^2_\nu = \left[{\sum_{i}^{N}\left(\frac{m_i-a}{\text{d}m_i}\right)^2}\right]/(N-1),\\
    \text{variability} \equiv \log10\left(\chi^2_\nu\right),
\end{split}
\end{equation}
where $N$ is the number of epochs, $m_i$ is the magnitude measured at the $i$th epoch, $a$ is the constant line of best fit through the datapoints and d$m_i$ is the error on the magnitude at the $i$th epoch. We emphasise that variability, as referred to in this paper, is the logarithm of the reduced $\chi^2$ as shown in Equation \ref{eqn:var}.

Under the hypothesis that the galaxy photometry is non-variable, the photometric data form a flat line as a function of date of observation. The reduced $\chi ^2$ tests this hypothesis, and provides a measure of how scattered the points are about the baseline, i.e. how variable the magnitudes of the galaxy are over the entire observation period. The result is a measure of the variability of the galaxies that depends on the magnitudes and the errors on the magnitudes. This means that dim galaxies with large uncertainties on the magnitude will have systematically underestimated variabilities compared to bright galaxies with small uncertainties, regardless of the actual variability of the source. To counteract this effect, we simulate variabilities and magnitudes of 10000 galaxies. The full details of the simulations are included in Appendix \ref{sec:sim}. The simulations allow us to match a `true' variability with the observed variability, at a given magnitude. Note that variability as described here is a dimensionless measure of of the scatter of points about a baseline, with variabilities $>1$ roughly indicating a variable source. 

For the purposes of this work, we decide to employ a magnitude cut when considering variabilities. This is because of the large disparity in distributions of magnitudes in AGN and star-forming galaxies, and that the variability calculation depends on the magnitude of the source. In effect, we want to examine the trends without the uncertainties that different magnitude ranges have on the calculated variabilities. Due to large overlap in the distributions, we choose $17<m\leq18$ to be the magnitude slice; note that results are not sensitive to the choice of magnitude slice.

{One additional benefit of employing a narrow apparent magnitude cut is that it reduces the variation in source `extendedness' on the sky. A potential source of non-physical variation in magnitudes is that ZTF psf-fit photometry depends on the on-sky extent of the galaxy. By restricting to a narrow apparent magnitude, we naturally reduce the variation in on-sky extent. Intrinsically fainter galaxies will be nearer than intrinsically brighter galaxies, but such fainter galaxies are also intrinsically smaller on average. In this way, the on-sky extent of our sample (as measured by the Petrosian radius) is quite narrow with a peak at $\sim$ 2 arcsecs, and the vast majority of galaxies less than 5 arcsecs. This on-sky extent is also largely consistent with stellar mass, with a slight tendency for more massive galaxies to appear larger. As we will discuss later, the effect of increased variation from large galaxies would act to reduce our results rather than explain them.}

\begin{figure}
    \centering
    \includegraphics[width=\columnwidth]{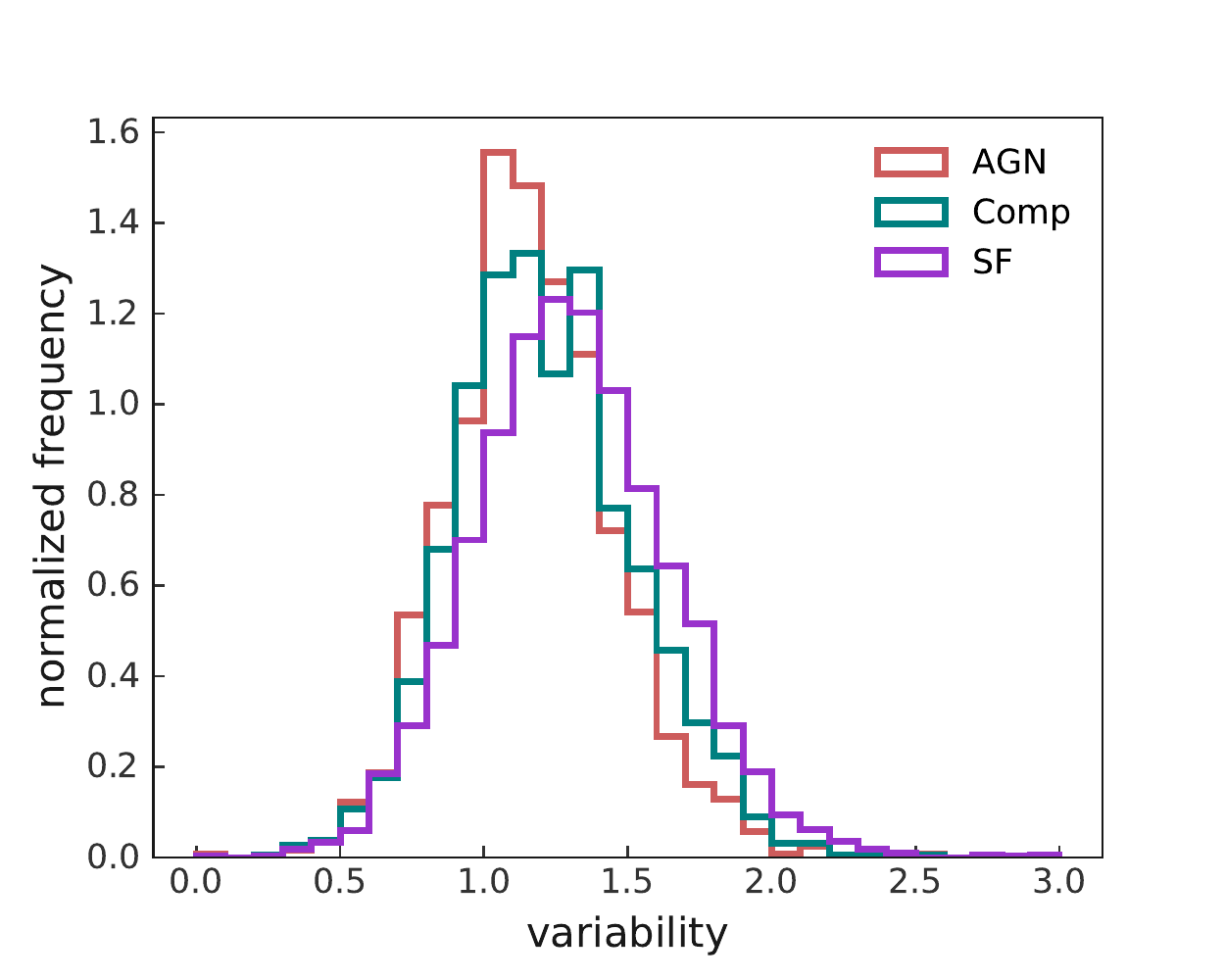}
    \caption{Variability distributions of AGN, star-forming galaxies, and composite galaxies. Here, distributions are normalised so that the area under each curve is 1.}
    \label{fig:varhist}
\end{figure}

The resulting distribution of variabilities for BPT diagram-defined AGN, composite galaxies, and star-forming galaxies are shown in Figure \ref{fig:varhist}. In this Figure, the classes are individually normalized to have an area under their curve of 1. The number of galaxies in each of these classes is 2858 (AGN), 3729 (composite) and 9990 (star-forming). We see that, unexpectedly, star-forming galaxies exhibit higher levels of variability than their counterparts with more active black holes via emission line diagnostics. By fitting Gaussians to these histograms, we can measure the means and standard deviations of each distribution; we find {$\sigma_\text{AGN} = 0.299\pm0.021$, $\sigma_\text{comp} = 0.298\pm0.026$ and $\sigma_\text{sf} = 0.341\pm0.012$, and $\mu_\text{AGN}=1.153\pm0.016$, $\mu_\text{comp}=1.182\pm0.011$ and $\mu_\text{sf}=1.300\pm0.010$}. We find slight differences in the widths of the distributions, but significant differences in the means, with $\mu_\text{sf}>\mu_\text{comp}>\mu_\text{AGN}$. This result will be expanded upon and discussed throughout the rest of the paper.

\subsection{Variability and galaxy properties}
\begin{figure}
    \centering
    \includegraphics[width=\columnwidth]{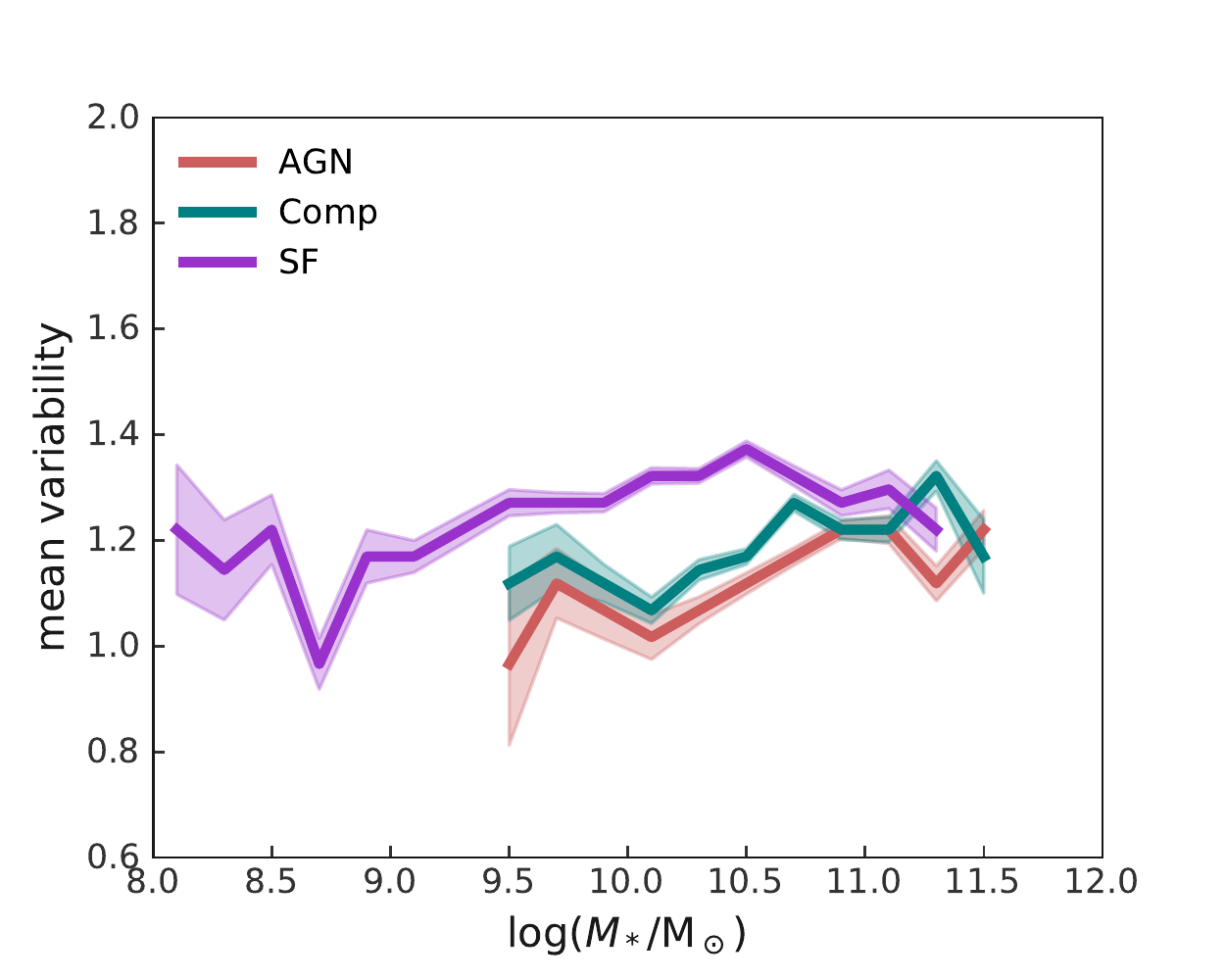}
    \caption{Mean variability binned by stellar mass, separated by emission type. Shaded regions indicate the standard deviation of the mean in that bin. Bins with fewer than 3 galaxies are omitted for clarity.}
    \label{fig:meanvarmass}
\end{figure}

We may now explore the relationships between variability and a variety of galactic properties. First, it must be stressed that many of these properties are highly inter-dependent and thus the exact causality of the changes in variability is difficult to identify. Nevertheless, in this section we attempt to present the data as is, and more in-depth discussion will take place in Section \ref{sec:discu}. In Figure \ref{fig:meanvarmass}, we plot the mean variability in each stellar mass bin separated by emission type as determined by the BPT diagram. The AGN and composite galaxies have very similar mean variabilities as a function of stellar mass, with both increasing from variability $\sim$ 1 at $M_* \sim 10^{9.5}$ M$_\odot$ to $\sim$ 1.2 at $M_* \sim 10^{11.5}$ M$_\odot$. In contrast, in all but the most massive bins ($M_*>10^{11} \text{M}_\odot$), the star-forming galaxies have mean variabilities that exceed that of AGN and composites. At $M_* = 10^{10} \text{M}_\odot$ the difference is 0.305 which is approximately 47 per cent of the width of the star-forming variability Gaussian, thus this is a significant deviation between the star-forming and AGN populations. Additionally, we see that the peak in mean variability occurs at $M_*\sim10^{10.5}\text{M}_\odot$ for star-forming galaxies and starts to decline at higher masses. In contrast, the AGN and composite mean variabilities appear roughly monotonically increasing with stellar mass.


\begin{figure}
    \centering
    \includegraphics[width=\columnwidth]{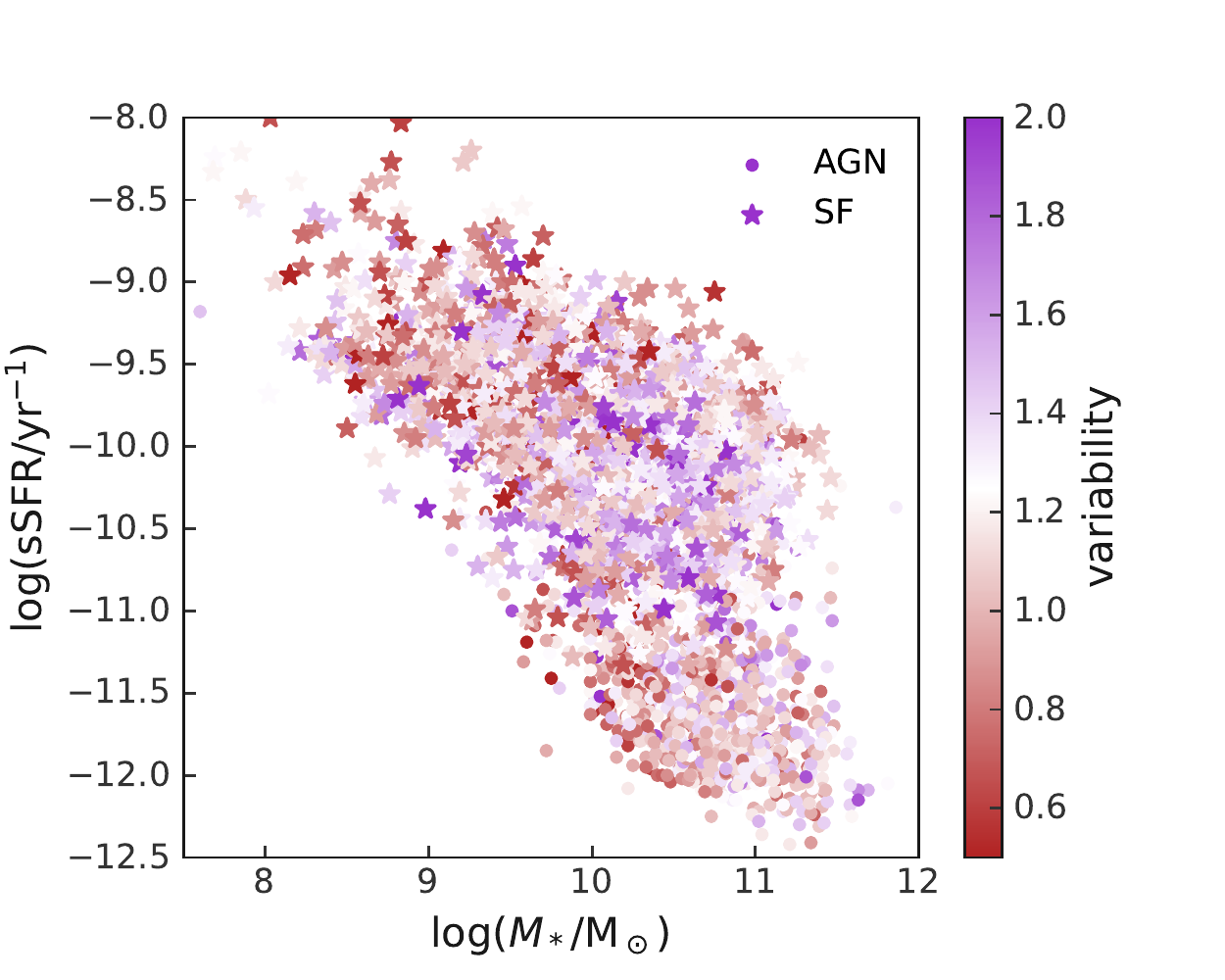}
    \caption{The relation between sSFR and $M_*$ of galaxies, separated by emission type (AGN [circle] and star-forming [star], composite galaxies are omitted for clarity), and colour-coded by variability.}
    \label{fig:ssfrscat}
\end{figure}

The interesting dependence of variability on stellar mass from Figure \ref{fig:meanvarmass} suggests an investigation into the variability as a function of stellar mass and specific star-formation rate (sSFR). The specific star-formation rate is the star-formation rate divided by the stellar mass, and gives an indication of how rapidly the galaxy is forming stars relative to its past average. A sSFR of $\log(\text{sSFR})=-11 \text{yr}^{-1}$ is commonly used as a demarcation between quenched and star-forming galaxies at z = 0, with star-forming galaxies having a higher sSFR \citep[eg.,][]{Wetzel2013, Cleland2021}. In Figure \ref{fig:ssfrscat}, we show the sSFR and stellar mass measurements for the AGN and star-forming classes and color each datapoint by their measured variability. We see that AGN (marked as circles) dominate the quenched region of the sSFR-$M_*$ plane, and that SF galaxies (marked by stars) are typically found above this region, as expected. Note that in this Figure, composite galaxies are omitted for clarity. Interestingly, by examining the figure, it appears that there is a band of high variability in the intermediate sSFR region, roughly $10^{-11} \text{yr}^{-1}<\text{sSFR}<10^{-10} \text{yr}^{-1}$. This is where the vast majority of the `purple' points from the high end of the variability color map lie. It also appears that both the high and low end of the sSFR distributions have lower variabilities than this middle region.  


\begin{figure*}
    \centering
    \includegraphics[width=\textwidth]{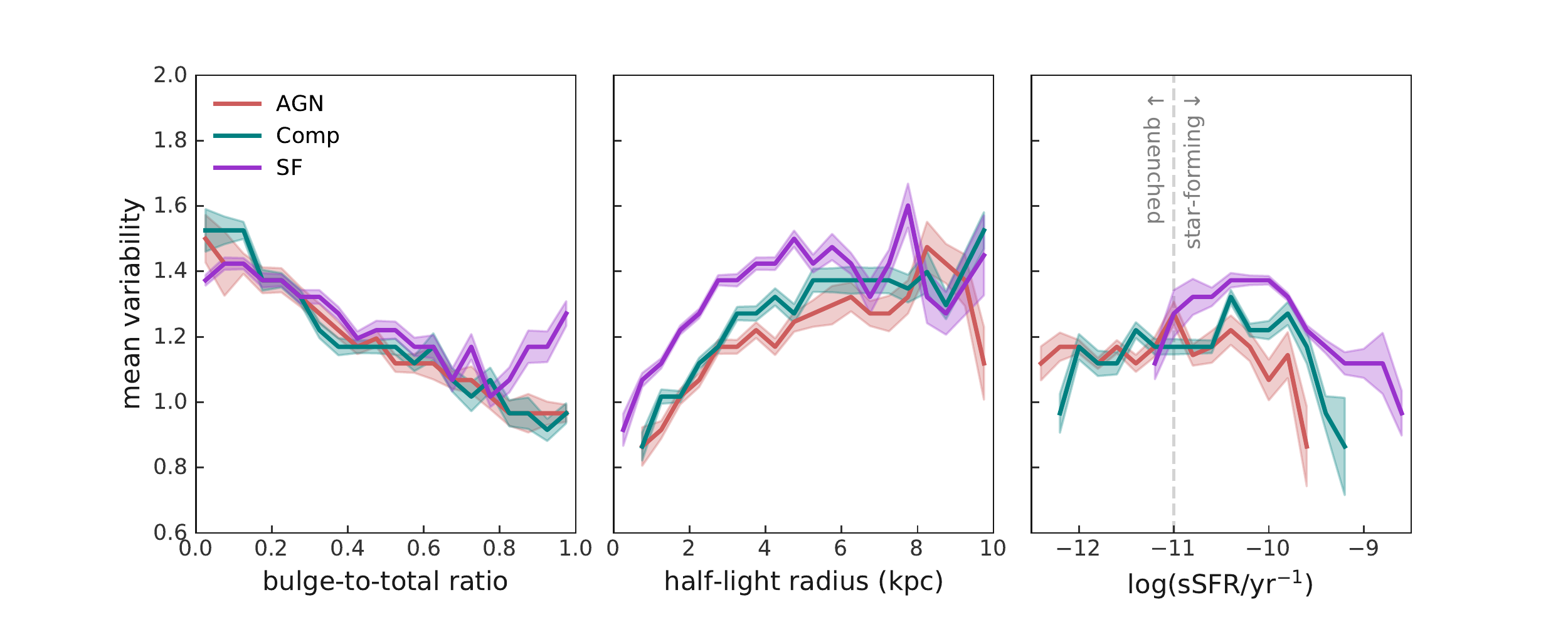}
    \caption{Mean variability binned by different quantities, separated by emission type. Shaded regions indicate the standard deviation of the mean in that bin. Bins with fewer than 3 galaxies are omitted for clarity. Left: mean variability against bulge-to-total ratio. Centre: mean variability against half-light radius. Right: mean variability against $\log(\text{sSFR}/\text{yr}^{-1})$.}
    \label{fig:btr}
\end{figure*}

In Figure \ref{fig:btr}, several trends emerge with respect to the mean variability binned by various galactic properties for the three galaxy classes. In the left panel, we plot the mean variability with respect to the bulge-to-total ratio ($B/T$) as measured by \citet{simard11} using bulge to disk decompositions. We see that for the most part, the mean variability is similar for each class and is steadily declining with the bulge-to-total ratio, deviating only at extreme values of $B/T$. However, this similarity of the galaxy classes does not hold when looking at the half-light radius ($R$) in the centre panel. We see that at almost all radii $R\lesssim 7$ kpc, star-forming galaxies have higher mean variabilities in each bin than composite galaxies and AGN. This is explained by the fact that at the same galaxy size, star-forming galaxies will typically have less massive bulges than AGN do, and thus will have a smaller $B/T$. As we see in the left panel of Figure \ref{fig:btr}, galaxies with smaller $B/T$ have higher mean variabilities, so the excess in mean variability for star-forming galaxies in the centre panel is accounted for. The right panel in Figure \ref{fig:btr} is similar to Figure \ref{fig:ssfrscat}, but better illustrates the trend we identified in which variability is highest in intermediate sSFR of star-forming galaxies. Interestingly, we also see stark declines in mean variability at $\log(\text{sSFR})\gtrsim-10 \text{yr}^{-1}$, indicating it is not sufficient to have high rates of star-formation to have high levels of variability.


\section{Discussion}\label{sec:discu}
\begin{figure*}
    \centering
    \includegraphics[width=\textwidth]{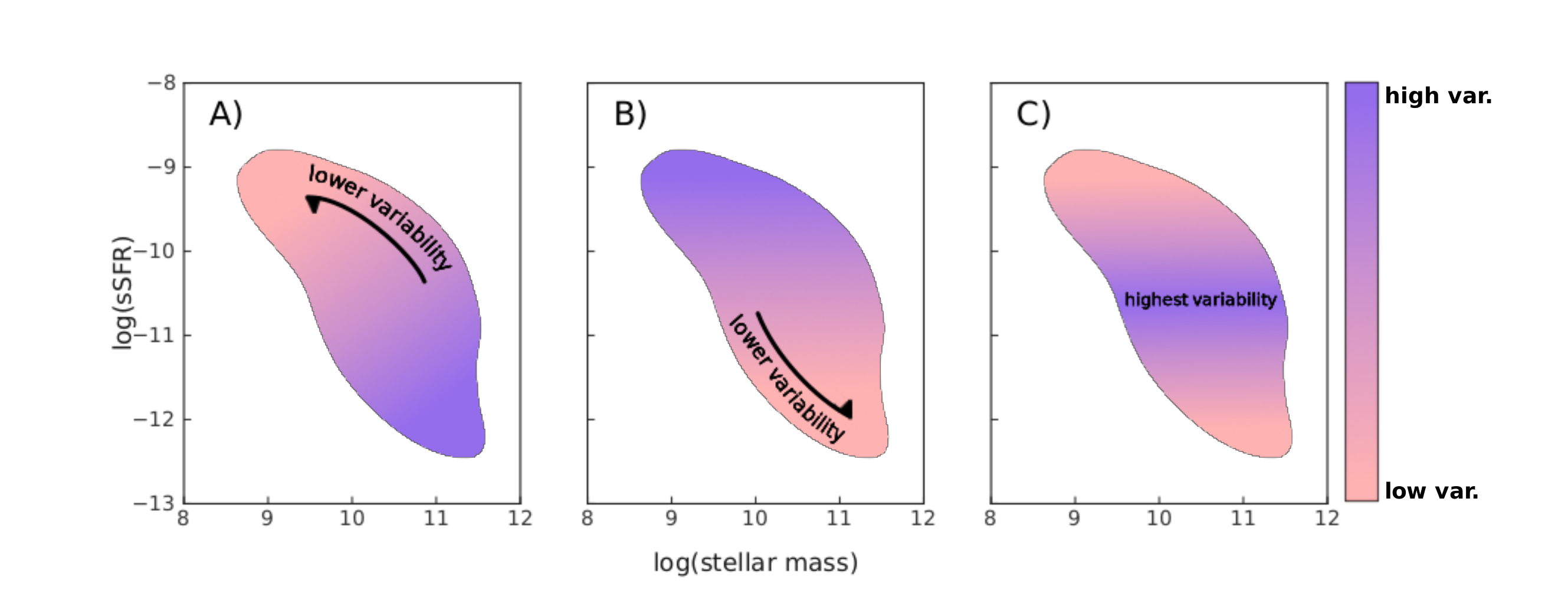}
    \caption{Three hypothetical scenarios that could possibly describe the behaviour of variability in sSFR-$M_*$ space. Scenario A) describes a situation where the BPT identifies all the AGN, and so variability is tightly linked to stellar mass and lack of star-formation. Scenario B) illustrates the idea of variability being dependent on gas content in a galaxy, so higher gas content implies higher SFR implies higher variability. Scenario C) is the situation we see in Figure \ref{fig:ssfrscat}, where a galaxy funnels gas into the centre after some time forming stars, exhibits variability, consumes the gas and quenches.}
    \label{fig:vardiagram}
\end{figure*}

\begin{figure}
    \centering
    \includegraphics[width=\columnwidth]{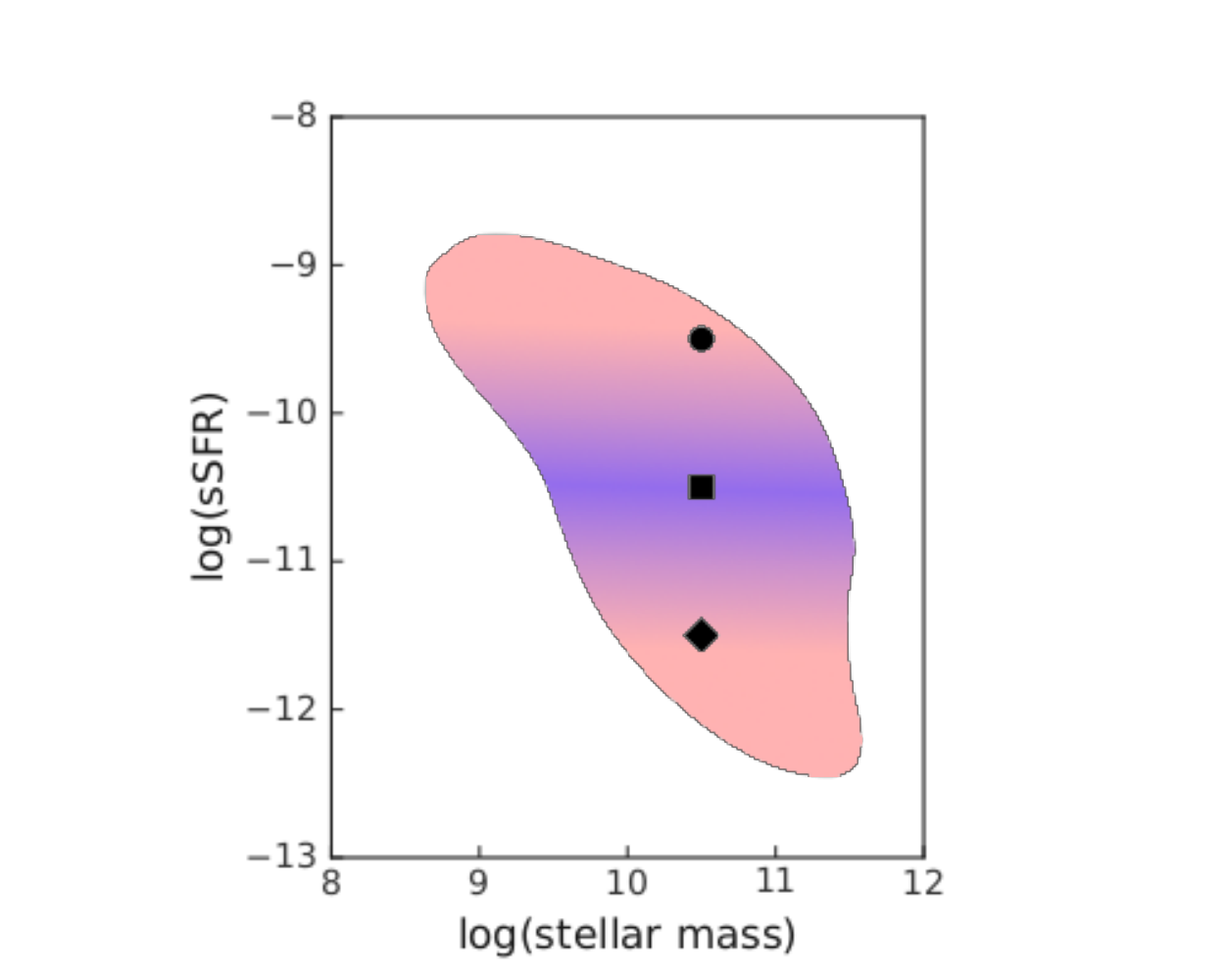}
    \caption{Scenario C) illustrated further; see text for full details.}
    \label{fig:diag2}
\end{figure}

We have found several empirical relations. We see that galaxies whose optical line emission is predominately the result of star-formation have higher variabilities (in the mean) than galaxies whose optical line emission is from AGN or a composite of AGN and star-formation. Further, we see that star-forming galaxies have higher variabilities at all stellar masses than composite or AGN. Interestingly, we showed that at fixed bulge-to-total ratio the mean variabilities of all three classes were similar, but at fixed half-light radius, star-forming galaxies had higher mean variabilities than seen in AGN or composite galaxies. Perhaps most interestingly, we found variability is strongly dependent on the galaxy location within the sSFR-$M_*$ plane. The star-forming galaxies have mean variability higher than either the AGN or composite galaxies at any fixed sSFR, but star-forming galaxies have a peak in their variability just `above' the quenched demarcation line.

\subsection{Source of the variability}
Until now, we have not discussed what actually causes the variability we see in these galaxies. There are at least three possibilities: 1. unaccounted-for errors in the photometric measurements, 2. intrinsic variability in the light from the dust/gas/stars in the galaxy, and 3. intrinsic variability caused by emission from the central black hole and associated accretion disk. We will look at each of these possibilities and the expectations of their signal.

The most obvious explanation for variability in photometric data is an unaccounted-for variation in the photometric measurements -- whether from inaccurate zeropoint calibration, unmatched aperture measurements (from seeing, etc) or a similar observational problem. While it is difficult to rule this out as a factor in all of the computed variabilities, we have taken a number of steps to minimize the possibility for error. For instance, we have principally examined the mean variability for wide classes of galaxies rather than the `extremes' of a particular distribution. We have limited our sample of galaxies to only one magnitude of apparent magnitude ($17<m\leq18$) to avoid the reliance on photometric error bars to be well represented across a wide dynamical range. Finally, if the mean variabilities we have measured were completely determined by unaccounted-for photometric errors, we would not expect these to show the dependence on physical galaxy properties that we have seen in the previous section. {The main way that a physical property could induce photometric errors would be through a correlation of the errors with the physical extent of the galaxy. The on-sky extent of a galaxy is likely to have a positive correlation with photometric errors. However, as we have discussed earlier, our narrow magnitude cut limits the correlation between the physical extent of the galaxy and the 'on-sky' extent. We have found the galaxies with higher stellar mass tend to have slightly higher on-sky extents on average. But, this would tend to induce increased errors as a function of stellar mass in the opposite manner of our results.} Given this, it is worth looking at the possible causes of intrinsic variability in the galaxy light. 

The next possibility is that the variability comes from the interstellar gas/dust or stars in the galaxy itself. The galaxies in this paper have half-light radii of $\sim$ 1 - 10 kpc, which would imply that the timescale for {\it complete} variability across the galaxy would have to be greater than $\sim$ 3000 - 30,000 years (assuming unimpeded light travel across the galaxy $\frac{r}{c}$). However, the variability we see is on the $\sim$ week to year timescale. Thus, the variability would have to be on smaller scales - individual stars or gas clouds. In this case, the individual stars and clouds would have to have widely and rapidly varying luminosities to avoid being washed away by the uncorrelated variations of other stars. This is clearly not possible for the normal variation of stars. The only possibility would be the large variation caused by supernovae or similar cataclysmic events. However, the variation we measure appears stochastic (Figure \ref{fig:lightcurve}) rather than the result of a particular rare event. {Although there have been recent suggestions that even supernova or stellar variation can mimic some of the AGN-like variation \citep{Burke2020}, these events are still expected to be rare.} As we are principally looking at the mean variability, we are not driven by these uncommon occurrences.

That leaves the possibility that the variation we measure is due to variable emission from the central black hole/AGN in each galaxy. The scale of central black holes and their accretion disks are thought to be $< 0.1$ pc which would imply light crossing times of $< 100$ days \citep{Hawkins2007, Morgan2010}. Although it is unlikely that global changes in the accretion disks are made on that timescale, it does indicate that large changes are possible on short timescales {possibly through thermal fluctuations of the UV-emitting region \citep{Kelly2009, Burke2021b}}. Indeed, direct observations of AGN have shown that $>$ order of magnitude changes are possible on hour to year timescales \citep{McHardy2006, Hickox2014}. Given the known variability of AGN, and the lack of an viable alternative source for the variability, in the rest of the paper will we consider what implications this variability has for our understanding of galaxy formation if it is due to AGN.

\subsection{Implications for galaxy formation if variability is from AGN}

Our finding that higher variability occurs at particular locations depending on galaxy type, stellar mass and specific star-formation rate may seem counter-intuitive to the current understanding of AGN physics. In this section, we attempt to illustrate a possible scenario for the physical mechanisms behind these results, and rule other possibilities out.

In Figure \ref{fig:vardiagram}, there are 3 plots illustrating possible physical scenarios which could be expected to occur in Figure \ref{fig:ssfrscat}, labelled A), B) and C).
These three scenarios are roughly: A) variability traces the appearance of emission line galaxies classed as AGNs, B) the variability traces the global sSFR, or C) variability traces those galaxies that are in the process of being most actively quenched. We will now discuss each in turn.

In scenario A), we illustrate the variability trend in sSFR-$M_*$ assuming that a BPT diagram correctly and completely identifies all the AGN in the galaxy sample. In this case, any star-forming galaxy will neither have an active black hole nor exhibit variability. Tracing these emission classes would result in the behavior shown in Figure \ref{fig:vardiagram} A). We see the highest variability would be found at the low-sSFR--high-mass end of the plot. Moving up the plot, variability decreases. In essence, scenario A) implies that variability is correlated with stellar mass, and anti-correlated with star-formation. While the assumption that star-forming galaxies have no active black hole is extreme, it is often treated as such when analysing star-forming galaxies.

In scenario B), we show the variability trend assuming that variability is fueled by global gas content in the galaxy, and thus is correlated with sSFR. This scenario would result in the behavior illustrated by Figure \ref{fig:vardiagram} B), which shows a region of high variability at high sSFR, with variability decreasing as sSFR decreases. While this complete correlation may seem extreme, it is a natural outcome of the often assumed tight relation between star-formation and black hole growth \citep[eg.,][]{Hopkins2006}.

However, neither scenario A) nor B) resemble what is seen in Figure \ref{fig:ssfrscat}. Figure \ref{fig:vardiagram} C) is intended to schematically represent what is actually happening; rather than finding high variability at the extremes of the sSFR-$M_*$ plot, we see a band of high variability at intermediate sSFR. We saw that a band of high variability occurred in the middle range of sSFR values, and so, the data rules out either of the extreme cases presented in A) and B). Assuming the variability traces AGN activity, then 1) there are active black holes in galaxies classified as star-forming via their emission line ratios and 2) there is not a direct instantaneous relation between the sSFR and the black hole growth rate. Scenario C) may imply a convolution of scenario A) and B).

Scenario C) can be described using the 3 points in Figure \ref{fig:diag2}; a circle, a square, and a diamond. The circle represents a galaxy in the early stages of its evolution, that has a high sSFR. Our data imply this galaxy has gas fuelling star-formation but relatively little gas feeding the black hole, thus, the galaxy shows the required low variability. The square represents a galaxy which now has gas feeding the black hole, and also has gas fuelling star-formation. The presence of gas in the centre of the galaxy provides the required variability. However, it would appear that due to AGN feedback and gas consumption, the galaxy's star-formation drops. So, the square galaxy has an intermediate sSFR and high variability. After some time, AGN feedback suppresses star-formation, thus the galaxy becomes quenched, as indicated by the diamond. Due to gas consumption, gas can no longer feed the black hole, and so the variability decreases, as required.

This scenario is consistent with other results throughout this work; in Figure \ref{fig:meanvarmass}, mean variability in SF galaxies peaks at $M_*\sim10^{10.5}$M$_\odot$, implying that as the bulge grows in a SF galaxy, variability increases until the point where the AGN `turns on' (as determined by the galaxy emission). Similarly, in Figure \ref{fig:btr}, we see that galaxies with underdeveloped bulges have higher variabilities compared to galaxies which have had longer to build their bulges. 
{In our possible scenarios, we have have not explicitly addressed the possibility that host galaxy dilution plays some role. For instance, two black holes of the same mass, accreting at the same rate, would likely have different contributions to the galaxy-wide variability if put in two galaxies with significantly different luminosities. It is very hard to rule out this effect playing a role, but we expect it would work to reduce the trends we see. That is, at a fixed stellar mass, star forming galaxies will have higher luminosities than AGN or composite galaxies -- because their more recent star formation leads to lower M/L ratios. So, we would expect host-dilution to muffle the signal in SF galaxies. Given we see the highest signal at fixed stellar mass in the SF galaxies, we expect this effect is not dominant. }

\section{Conclusions} \label{sec:conc}

We have combined photometric lightcurve data from the Zwicky Transient Facility with spectroscopic data from the Sloan Digital Sky Survey to examine the variability of $\sim$ 16500 galaxies at z $<$ 0.2. Using the value-added measurements of SDSS, we examine this variability as a function of stellar mass, specific star-formation rate, galaxy class as determined by emission line ratios, effective radii and bulge to total ratios. Our main findings can be summarised as: 

\begin{itemize}
    \item The photometric variability of star-forming galaxies (with their class determined by their location in the BPT diagram) is higher than the variability of either the composite or AGN-dominated galaxies. Restricting to a narrow slice in apparent magnitude (17 $<$ $m$ $\leq$ 18), we find the median variability of star-forming galaxies is 1.289, while it is 1.150 for AGN and 1.190 for composite galaxies.
    \item The elevated variability in star-forming galaxies compared to AGN or composite galaxies is also seen at fixed stellar mass for all but the highest ranges ($>$ 10$^{11}$ M$_\odot$).
    \item The highest variabilities occur in a well defined region of the sSFR-$M_*$ plane -- in a band with -11 $>$ log(sSFR/yr$^{-1}$) $>$ -10. In this region, we again see that star-forming galaxies have the highest variabilities of the three galaxy classes.
    \item The differences in mean variabilties of the galaxy classes largely go away when compared at fixed bulge-to-total ratio (for $B/T<0.8$), but remain for small half-light ratios $<$ 6 kpc.
\end{itemize}

We argue that the best explanation, based on timing arguments, for the observed behavior of mean variabilities is that the variability is driven by AGN activity. Under this assumption, we argue this provides evidence that central supermassive black holes exist in star-forming galaxies, including in low mass galaxies. We also argue that the appearance of the mean variabilities rules out an exact correlation between instantaneous star-formation and black hole growth. Finally, we argue this enhanced variability is evidence that AGN feedback is shutting down the star-formation in actively star-forming galaxies.  
Taken as a whole, these results indicate that emission in the form of star-formation may obscure a low-mass AGN, and hide the dramatic effects the AGN are having on galaxy evolution.

{In our work, we have attempted to measure the variation of a well-selected sample of galaxies in a `source-agnostic' way. Given our current explanation that this variation arises chiefly from AGN activity, in future work we will explore further measures to quantify the AGN. We plan to use both the $g$- and $r$-bands to measure the variability with other techniques such as by quantifying the RMS of the intrinsic source variability \citep{Shen2019}, and comparing to expected summary statistics of stochastic variations \citep{Kelly2009, Kozlowski2016}. We also expect to tie the requirements for the length of light curves more directly to possible time-frames which maximize the coverage of relevant BH timescales \citep[eg.,][]{Burke2021b}.}




\section*{Acknowledgements}
The authors thank the referee for a careful and helpful report which has improved the paper.
CC acknowledges the support of the School of Physics and Astronomy at the University of Birmingham. SLM acknowledges support from the Science and Technology Facilities Council through grant number ST/N021702/1.

Funding for the SDSS and SDSS-II has been provided by the Alfred P. Sloan Foundation, the Participating Institutions, the National Science Foundation, the U.S. Department of Energy, the National Aeronautics and Space Administration, the Japanese Monbukagakusho, the Max Planck Society, and the Higher Education Funding Council for England. The SDSS Web Site is http://www.sdss.org/.

The SDSS is managed by the Astrophysical Research Consortium for the Participating Institutions. The Participating Institutions are the American Museum of Natural History, Astrophysical Institute Potsdam, University of Basel, University of Cambridge, Case Western Reserve University, University of Chicago, Drexel University, Fermilab, the Institute for Advanced Study, the Japan Participation Group, Johns Hopkins University, the Joint Institute for Nuclear Astrophysics, the Kavli Institute for Particle Astrophysics and Cosmology, the Korean Scientist Group, the Chinese Academy of Sciences (LAMOST), Los Alamos National Laboratory, the Max-Planck-Institute for Astronomy (MPIA), the Max-Planck-Institute for Astrophysics (MPA), New Mexico State University, Ohio State University, University of Pittsburgh, University of Portsmouth, Princeton University, the United States Naval Observatory, and the University of Washington.

Based on observations obtained with the Samuel Oschin 48-inch Telescope at the Palomar Observatory as part of the Zwicky Transient Facility project. ZTF is supported by the National Science Foundation under Grant No. AST-1440341 and a collaboration including Caltech, IPAC, the Weizmann Institute for Science, the Oskar Klein Center at Stockholm University, the University of Maryland, the University of Washington, Deutsches Elektronen-Synchrotron and Humboldt University, Los Alamos National Laboratories, the TANGO Consortium of Taiwan, the University of Wisconsin at Milwaukee, and Lawrence Berkeley National Laboratories. Operations are conducted by COO, IPAC, and UW. 

\section*{Data Availability}

The data used in this work are available at the URLs linked where they are discussed in the article.



\bibliographystyle{mnras}
\bibliography{bib} 

\begin{thebibliography}{}
\makeatletter
\relax
\def\mn@urlcharsother{\let\do\@makeother \do\$\do\&\do\#\do\^\do\_\do\%\do\~}
\def\mn@doi{\begingroup\mn@urlcharsother \@ifnextchar [ {\mn@doi@}
  {\mn@doi@[]}}
\def\mn@doi@[#1]#2{\def\@tempa{#1}\ifx\@tempa\@empty \href
  {http://dx.doi.org/#2} {doi:#2}\else \href {http://dx.doi.org/#2} {#1}\fi
  \endgroup}
\def\mn@eprint#1#2{\mn@eprint@#1:#2::\@nil}
\def\mn@eprint@arXiv#1{\href {http://arxiv.org/abs/#1} {{\tt arXiv:#1}}}
\def\mn@eprint@dblp#1{\href {http://dblp.uni-trier.de/rec/bibtex/#1.xml}
  {dblp:#1}}
\def\mn@eprint@#1:#2:#3:#4\@nil{\def\@tempa {#1}\def\@tempb {#2}\def\@tempc
  {#3}\ifx \@tempc \@empty \let \@tempc \@tempb \let \@tempb \@tempa \fi \ifx
  \@tempb \@empty \def\@tempb {arXiv}\fi \@ifundefined
  {mn@eprint@\@tempb}{\@tempb:\@tempc}{\expandafter \expandafter \csname
  mn@eprint@\@tempb\endcsname \expandafter{\@tempc}}}

\bibitem[\protect\citeauthoryear{{Abazajian} et~al.,}{{Abazajian}
  et~al.}{2009}]{abazajian09}
{Abazajian} K.~N.,  et~al., 2009, \mn@doi [\apjs]
  {10.1088/0067-0049/182/2/543}, \href
  {https://ui.adsabs.harvard.edu/abs/2009ApJS..182..543A} {182, 543}

\bibitem[\protect\citeauthoryear{{Agostino} et~al.,}{{Agostino}
  et~al.}{2021}]{Agostino2021}
{Agostino} C.~J.,  et~al., 2021, arXiv e-prints, \href
  {https://ui.adsabs.harvard.edu/abs/2021arXiv210807812A} {p. arXiv:2108.07812}

\bibitem[\protect\citeauthoryear{{Baldassare}, {Geha}  \&
  {Greene}}{{Baldassare} et~al.}{2018}]{baldassare18}
{Baldassare} V.~F.,  {Geha} M.,   {Greene} J.,  2018, \mn@doi [\apj]
  {10.3847/1538-4357/aae6cf}, \href
  {https://ui.adsabs.harvard.edu/abs/2018ApJ...868..152B} {868, 152}

\bibitem[\protect\citeauthoryear{{Baldassare}, {Geha}  \&
  {Greene}}{{Baldassare} et~al.}{2020}]{baldassare20}
{Baldassare} V.~F.,  {Geha} M.,   {Greene} J.,  2020, \mn@doi [\apj]
  {10.3847/1538-4357/ab8936}, \href
  {https://ui.adsabs.harvard.edu/abs/2020ApJ...896...10B} {896, 10}

\bibitem[\protect\citeauthoryear{{Baldwin}, {Phillips}  \&
  {Terlevich}}{{Baldwin} et~al.}{1981}]{baldwin81}
{Baldwin} J.~A.,  {Phillips} M.~M.,   {Terlevich} R.,  1981, \mn@doi [\pasp]
  {10.1086/130766}, \href
  {https://ui.adsabs.harvard.edu/abs/1981PASP...93....5B} {93, 5}

\bibitem[\protect\citeauthoryear{{Bellm}}{{Bellm}}{2014}]{bellm14}
{Bellm} E.,  2014, in {Wozniak} P.~R.,  {Graham} M.~J.,  {Mahabal} A.~A.,
  {Seaman} R.,  eds, The Third Hot-wiring the Transient Universe Workshop. pp
  27--33 (\mn@eprint {arXiv} {1410.8185})

\bibitem[\protect\citeauthoryear{{Bennett} et~al.,}{{Bennett}
  et~al.}{2013}]{bennett13}
{Bennett} C.~L.,  et~al., 2013, \mn@doi [\apjs] {10.1088/0067-0049/208/2/20},
  \href {https://ui.adsabs.harvard.edu/abs/2013ApJS..208...20B} {208, 20}

\bibitem[\protect\citeauthoryear{{Bland-Hawthorn}, {Maloney}, {Sutherland}  \&
  {Madsen}}{{Bland-Hawthorn} et~al.}{2013}]{blandhawthorn13}
{Bland-Hawthorn} J.,  {Maloney} P.~R.,  {Sutherland} R.~S.,   {Madsen} G.~J.,
  2013, \mn@doi [\apj] {10.1088/0004-637X/778/1/58}, \href
  {https://ui.adsabs.harvard.edu/abs/2013ApJ...778...58B} {778, 58}

\bibitem[\protect\citeauthoryear{{Bower}, {Benson}, {Malbon}, {Helly}, {Frenk},
  {Baugh}, {Cole}  \& {Lacey}}{{Bower} et~al.}{2006}]{Bower2006}
{Bower} R.~G.,  {Benson} A.~J.,  {Malbon} R.,  {Helly} J.~C.,  {Frenk} C.~S.,
  {Baugh} C.~M.,  {Cole} S.,   {Lacey} C.~G.,  2006, \mn@doi [\mnras]
  {10.1111/j.1365-2966.2006.10519.x}, \href
  {https://ui.adsabs.harvard.edu/abs/2006MNRAS.370..645B} {370, 645}

\bibitem[\protect\citeauthoryear{{Brinchmann}, {Charlot}, {White}, {Tremonti},
  {Kauffmann}, {Heckman}  \& {Brinkmann}}{{Brinchmann}
  et~al.}{2004}]{brinchmann04}
{Brinchmann} J.,  {Charlot} S.,  {White} S.~D.~M.,  {Tremonti} C.,  {Kauffmann}
  G.,  {Heckman} T.,   {Brinkmann} J.,  2004, \mn@doi [\mnras]
  {10.1111/j.1365-2966.2004.07881.x}, \href
  {https://ui.adsabs.harvard.edu/abs/2004MNRAS.351.1151B} {351, 1151}

\bibitem[\protect\citeauthoryear{{Burke} et~al.,}{{Burke}
  et~al.}{2020}]{Burke2020}
{Burke} C.~J.,  et~al., 2020, \mn@doi [\apjl] {10.3847/2041-8213/ab88de}, \href
  {https://ui.adsabs.harvard.edu/abs/2020ApJ...894L...5B} {894, L5}

\bibitem[\protect\citeauthoryear{{Burke} et~al.}{{Burke}
  et~al.}{2021a}]{Burke2021}
{Burke} C.~J.,  et~al., 2021a, arXiv e-prints, \href
  {https://ui.adsabs.harvard.edu/abs/2021arXiv211103079B} {p. arXiv:2111.03079}

\bibitem[\protect\citeauthoryear{{Burke} et~al.,}{{Burke}
  et~al.}{2021b}]{Burke2021b}
{Burke} C.~J.,  et~al., 2021b, \mn@doi [Science] {10.1126/science.abg9933},
  \href {https://ui.adsabs.harvard.edu/abs/2021Sci...373..789B} {373, 789}

\bibitem[\protect\citeauthoryear{{Cann}, {Satyapal}, {Abel}, {Blecha},
  {Mushotzky}, {Reynolds}  \& {Secrest}}{{Cann} et~al.}{2019}]{Cann2019}
{Cann} J.~M.,  {Satyapal} S.,  {Abel} N.~P.,  {Blecha} L.,  {Mushotzky} R.~F.,
  {Reynolds} C.~S.,   {Secrest} N.~J.,  2019, \mn@doi [\apjl]
  {10.3847/2041-8213/aaf88d}, \href
  {https://ui.adsabs.harvard.edu/abs/2019ApJ...870L...2C} {870, L2}

\bibitem[\protect\citeauthoryear{{Caplar}, {Lilly}  \& {Trakhtenbrot}}{{Caplar}
  et~al.}{2017}]{caplar17}
{Caplar} N.,  {Lilly} S.~J.,   {Trakhtenbrot} B.,  2017, \mn@doi [\apj]
  {10.3847/1538-4357/834/2/111}, \href
  {https://ui.adsabs.harvard.edu/abs/2017ApJ...834..111C} {834, 111}

\bibitem[\protect\citeauthoryear{{Cartier} et~al.,}{{Cartier}
  et~al.}{2015}]{cartier15}
{Cartier} R.,  et~al., 2015, \mn@doi [\apj] {10.1088/0004-637X/810/2/164},
  \href {https://ui.adsabs.harvard.edu/abs/2015ApJ...810..164C} {810, 164}

\bibitem[\protect\citeauthoryear{{Cleland} \& {McGee}}{{Cleland} \&
  {McGee}}{2021}]{Cleland2021}
{Cleland} C.,  {McGee} S.~L.,  2021, \mn@doi [\mnras] {10.1093/mnras/staa3267},
  \href {https://ui.adsabs.harvard.edu/abs/2021MNRAS.500..590C} {500, 590}

\bibitem[\protect\citeauthoryear{{Croton} et~al.,}{{Croton}
  et~al.}{2006}]{Croton2006}
{Croton} D.~J.,  et~al., 2006, \mn@doi [\mnras]
  {10.1111/j.1365-2966.2005.09675.x}, \href
  {https://ui.adsabs.harvard.edu/abs/2006MNRAS.365...11C} {365, 11}

\bibitem[\protect\citeauthoryear{{Di Matteo}, {Springel}  \& {Hernquist}}{{Di
  Matteo} et~al.}{2005}]{dimatteo05}
{Di Matteo} T.,  {Springel} V.,   {Hernquist} L.,  2005, \mn@doi [\nat]
  {10.1038/nature03335}, \href
  {https://ui.adsabs.harvard.edu/abs/2005Natur.433..604D} {433, 604}

\bibitem[\protect\citeauthoryear{{Fabian}}{{Fabian}}{2012}]{Fabian2012}
{Fabian} A.~C.,  2012, \mn@doi [\araa] {10.1146/annurev-astro-081811-125521},
  \href {https://ui.adsabs.harvard.edu/abs/2012ARA&A..50..455F} {50, 455}

\bibitem[\protect\citeauthoryear{{Hawkins}}{{Hawkins}}{2007}]{Hawkins2007}
{Hawkins} M.~R.~S.,  2007, \mn@doi [\aap] {10.1051/0004-6361:20066283}, \href
  {https://ui.adsabs.harvard.edu/abs/2007A&A...462..581H} {462, 581}

\bibitem[\protect\citeauthoryear{{Hickox}, {Mullaney}, {Alexander}, {Chen},
  {Civano}, {Goulding}  \& {Hainline}}{{Hickox} et~al.}{2014}]{Hickox2014}
{Hickox} R.~C.,  {Mullaney} J.~R.,  {Alexander} D.~M.,  {Chen} C.-T.~J.,
  {Civano} F.~M.,  {Goulding} A.~D.,   {Hainline} K.~N.,  2014, \mn@doi [\apj]
  {10.1088/0004-637X/782/1/9}, \href
  {https://ui.adsabs.harvard.edu/abs/2014ApJ...782....9H} {782, 9}

\bibitem[\protect\citeauthoryear{{Hopkins}, {Hernquist}, {Cox}, {Di Matteo},
  {Robertson}  \& {Springel}}{{Hopkins} et~al.}{2006}]{Hopkins2006}
{Hopkins} P.~F.,  {Hernquist} L.,  {Cox} T.~J.,  {Di Matteo} T.,  {Robertson}
  B.,   {Springel} V.,  2006, \mn@doi [\apjs] {10.1086/499298}, \href
  {https://ui.adsabs.harvard.edu/abs/2006ApJS..163....1H} {163, 1}

\bibitem[\protect\citeauthoryear{{Kauffmann} et~al.,}{{Kauffmann}
  et~al.}{2003}]{kauffmann03}
{Kauffmann} G.,  et~al., 2003, \mn@doi [\mnras]
  {10.1111/j.1365-2966.2003.07154.x}, \href
  {https://ui.adsabs.harvard.edu/abs/2003MNRAS.346.1055K} {346, 1055}

\bibitem[\protect\citeauthoryear{{Kelly}, {Bechtold}  \&
  {Siemiginowska}}{{Kelly} et~al.}{2009}]{Kelly2009}
{Kelly} B.~C.,  {Bechtold} J.,   {Siemiginowska} A.,  2009, \mn@doi [\apj]
  {10.1088/0004-637X/698/1/895}, \href
  {https://ui.adsabs.harvard.edu/abs/2009ApJ...698..895K} {698, 895}

\bibitem[\protect\citeauthoryear{{Kewley}, {Groves}, {Kauffmann}  \&
  {Heckman}}{{Kewley} et~al.}{2006}]{kewley06}
{Kewley} L.~J.,  {Groves} B.,  {Kauffmann} G.,   {Heckman} T.,  2006, \mn@doi
  [\mnras] {10.1111/j.1365-2966.2006.10859.x}, \href
  {https://ui.adsabs.harvard.edu/abs/2006MNRAS.372..961K} {372, 961}

\bibitem[\protect\citeauthoryear{{Kormendy} \& {Ho}}{{Kormendy} \&
  {Ho}}{2013}]{Kormendy2013}
{Kormendy} J.,  {Ho} L.~C.,  2013, \mn@doi [\araa]
  {10.1146/annurev-astro-082708-101811}, \href
  {https://ui.adsabs.harvard.edu/abs/2013ARA&A..51..511K} {51, 511}

\bibitem[\protect\citeauthoryear{{Koz{\l}owski}}{{Koz{\l}owski}}{2016}]{Kozlowski2016}
{Koz{\l}owski} S.,  2016, \mn@doi [\apj] {10.3847/0004-637X/826/2/118}, \href
  {https://ui.adsabs.harvard.edu/abs/2016ApJ...826..118K} {826, 118}

\bibitem[\protect\citeauthoryear{{Kroupa}}{{Kroupa}}{2001}]{kroupa01}
{Kroupa} P.,  2001, \mn@doi [\mnras] {10.1046/j.1365-8711.2001.04022.x}, \href
  {https://ui.adsabs.harvard.edu/abs/2001MNRAS.322..231K} {322, 231}

\bibitem[\protect\citeauthoryear{{Magorrian} et~al.,}{{Magorrian}
  et~al.}{1998}]{Magorrian1998}
{Magorrian} J.,  et~al., 1998, \mn@doi [\aj] {10.1086/300353}, \href
  {https://ui.adsabs.harvard.edu/abs/1998AJ....115.2285M} {115, 2285}

\bibitem[\protect\citeauthoryear{{Masci} et~al.,}{{Masci}
  et~al.}{2019}]{masci19}
{Masci} F.~J.,  et~al., 2019, \mn@doi [\pasp] {10.1088/1538-3873/aae8ac}, \href
  {https://ui.adsabs.harvard.edu/abs/2019PASP..131a8003M} {131, 018003}

\bibitem[\protect\citeauthoryear{{McConnell} \& {Ma}}{{McConnell} \&
  {Ma}}{2013}]{Mcconnell2013}
{McConnell} N.~J.,  {Ma} C.-P.,  2013, \mn@doi [\apj]
  {10.1088/0004-637X/764/2/184}, \href
  {https://ui.adsabs.harvard.edu/abs/2013ApJ...764..184M} {764, 184}

\bibitem[\protect\citeauthoryear{{McHardy}, {Koerding}, {Knigge}, {Uttley}  \&
  {Fender}}{{McHardy} et~al.}{2006}]{McHardy2006}
{McHardy} I.~M.,  {Koerding} E.,  {Knigge} C.,  {Uttley} P.,   {Fender} R.~P.,
  2006, \mn@doi [\nat] {10.1038/nature05389}, \href
  {https://ui.adsabs.harvard.edu/abs/2006Natur.444..730M} {444, 730}

\bibitem[\protect\citeauthoryear{{Morgan}, {Kochanek}, {Morgan}  \&
  {Falco}}{{Morgan} et~al.}{2010}]{Morgan2010}
{Morgan} C.~W.,  {Kochanek} C.~S.,  {Morgan} N.~D.,   {Falco} E.~E.,  2010,
  \mn@doi [\apj] {10.1088/0004-637X/712/2/1129}, \href
  {https://ui.adsabs.harvard.edu/abs/2010ApJ...712.1129M} {712, 1129}

\bibitem[\protect\citeauthoryear{{Novak}, {Ostriker}  \& {Ciotti}}{{Novak}
  et~al.}{2011}]{novak11}
{Novak} G.~S.,  {Ostriker} J.~P.,   {Ciotti} L.,  2011, \mn@doi [\apj]
  {10.1088/0004-637X/737/1/26}, \href
  {https://ui.adsabs.harvard.edu/abs/2011ApJ...737...26N} {737, 26}

\bibitem[\protect\citeauthoryear{{Omand}, {Balogh}  \& {Poggianti}}{{Omand}
  et~al.}{2015}]{omand15}
{Omand} C.~M.~B.,  {Balogh} M.~L.,   {Poggianti} B.~M.,  2015, VizieR Online
  Data Catalog, \href {https://ui.adsabs.harvard.edu/abs/2015yCat..74400843O}
  {p. J/MNRAS/440/843}

\bibitem[\protect\citeauthoryear{{Peng} et~al.}{{Peng} et~al.}{2010}]{peng10}
{Peng} Y.-j.,  et~al., 2010, \mn@doi [\apj] {10.1088/0004-637X/721/1/193},
  \href {https://ui.adsabs.harvard.edu/abs/2010ApJ...721..193P} {721, 193}

\bibitem[\protect\citeauthoryear{{Peng}, {Maiolino}  \& {Cochrane}}{{Peng}
  et~al.}{2015}]{Peng2015}
{Peng} Y.,  {Maiolino} R.,   {Cochrane} R.,  2015, \mn@doi [\nat]
  {10.1038/nature14439}, \href
  {https://ui.adsabs.harvard.edu/abs/2015Natur.521..192P} {521, 192}

\bibitem[\protect\citeauthoryear{{Reines} \& {Volonteri}}{{Reines} \&
  {Volonteri}}{2015}]{Reines2015}
{Reines} A.~E.,  {Volonteri} M.,  2015, \mn@doi [\apj]
  {10.1088/0004-637X/813/2/82}, \href
  {https://ui.adsabs.harvard.edu/abs/2015ApJ...813...82R} {813, 82}

\bibitem[\protect\citeauthoryear{{Sartori}, {Schawinski}, {Trakhtenbrot},
  {Caplar}, {Treister}, {Koss}, {Urry}  \& {Zhang}}{{Sartori}
  et~al.}{2018}]{Sartori2018}
{Sartori} L.~F.,  {Schawinski} K.,  {Trakhtenbrot} B.,  {Caplar} N.,
  {Treister} E.,  {Koss} M.~J.,  {Urry} C.~M.,   {Zhang} C.~E.,  2018, \mn@doi
  [\mnras] {10.1093/mnrasl/sly025}, \href
  {https://ui.adsabs.harvard.edu/abs/2018MNRAS.476L..34S} {476, L34}

\bibitem[\protect\citeauthoryear{{Shen} et~al.,}{{Shen}
  et~al.}{2019}]{Shen2019}
{Shen} Y.,  et~al., 2019, \mn@doi [\apjs] {10.3847/1538-4365/ab074f}, \href
  {https://ui.adsabs.harvard.edu/abs/2019ApJS..241...34S} {241, 34}

\bibitem[\protect\citeauthoryear{{Simard}, {Mendel}, {Patton}, {Ellison}  \&
  {McConnachie}}{{Simard} et~al.}{2011}]{simard11}
{Simard} L.,  {Mendel} J.~T.,  {Patton} D.~R.,  {Ellison} S.~L.,
  {McConnachie} A.~W.,  2011, \mn@doi [\apjs] {10.1088/0067-0049/196/1/11},
  \href {https://ui.adsabs.harvard.edu/abs/2011ApJS..196...11S} {196, 11}

\bibitem[\protect\citeauthoryear{{Smith}, {Mushotzky}, {Boyd}, {Malkan},
  {Howell}  \& {Gelino}}{{Smith} et~al.}{2018}]{smith18}
{Smith} K.~L.,  {Mushotzky} R.~F.,  {Boyd} P.~T.,  {Malkan} M.,  {Howell}
  S.~B.,   {Gelino} D.~M.,  2018, \mn@doi [\apj] {10.3847/1538-4357/aab88d},
  \href {https://ui.adsabs.harvard.edu/abs/2018ApJ...857..141S} {857, 141}

\bibitem[\protect\citeauthoryear{{Ward} et~al.,}{{Ward}
  et~al.}{2021}]{ward2021}
{Ward} C.,  et~al., 2021, arXiv e-prints, \href
  {https://ui.adsabs.harvard.edu/abs/2021arXiv211013098W} {p. arXiv:2110.13098}

\bibitem[\protect\citeauthoryear{{Wetzel}, {Tinker}, {Conroy}  \& {van den
  Bosch}}{{Wetzel} et~al.}{2013}]{Wetzel2013}
{Wetzel} A.~R.,  {Tinker} J.~L.,  {Conroy} C.,   {van den Bosch} F.~C.,  2013,
  \mn@doi [\mnras] {10.1093/mnras/stt469}, \href
  {https://ui.adsabs.harvard.edu/abs/2013MNRAS.432..336W} {432, 336}

\bibitem[\protect\citeauthoryear{{Zibecchi}, {Andruchow}, {Cellone}  \&
  {Carpintero}}{{Zibecchi} et~al.}{2020}]{zibecchi20}
{Zibecchi} L.,  {Andruchow} I.,  {Cellone} S.~A.,   {Carpintero} D.~D.,  2020,
  \mn@doi [\mnras] {10.1093/mnras/staa2544}, \href
  {https://ui.adsabs.harvard.edu/abs/2020MNRAS.498.3013Z} {498, 3013}

\makeatother
\end{thebibliography}




\appendix

\section{Simulations of recovered variability}\label{sec:sim}

\begin{figure}
    \centering
    \includegraphics[width=\columnwidth]{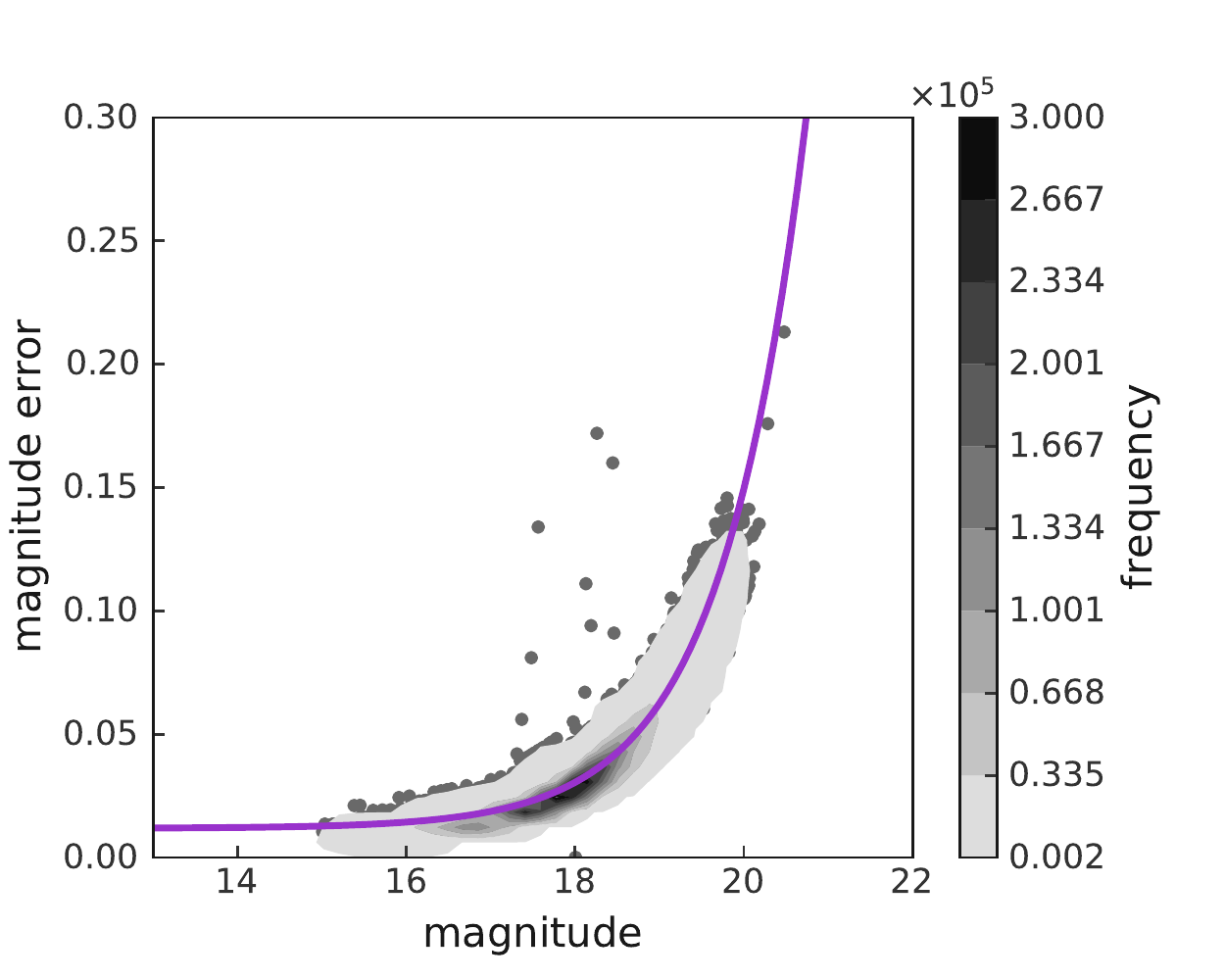}
    \caption{Magnitude against magnitude uncertainty for each epoch in the ZTF dataset. For plotting purposes, the number of datapoints has been reduced from $10^5$ to $10^4$.}
    \label{fig:magerrs}
\end{figure}

These simulations are motivated by the fact that without them, low variability is biased towards bright objects with small uncertainties. To account for this effect, without discarding the uncertainties entirely, we begin with a sample of 10,000 galaxies to which we randomly assign a variability and a magnitude. Both the variability and magnitudes are taken independently from uniform distributions. For clarity, this variability will be referred to as the `true' variability. We reconstruct a lightcurve using the true variability and some Gaussian scatter. In order to reconstruct the lightcurve from the variability, an average value for $\text{d}m$ must be assumed (see Equation \ref{eqn:var}). This value is taken from the best fit line through the magnitude uncertainties with respect to the magnitudes, as shown in Figure \ref{fig:magerrs}. 

When the simulated observations are in place, we assign a different uncertainty on the magnitude. This uncertainty is drawn from a sample of uncertainties which is based on the real uncertainties in the data for a given magnitude bin. We now recalculate the variability in the exact same way; this variability may be thought of as the `reco' variability, analogous to the observed variability in the real data. This allows for a direct comparison between the reco variability and the true variability, at a given magnitude. The resulting scatter acts as an uncertainty on the true variability.

\begin{figure*}
    \centering
    \includegraphics[width=\textwidth]{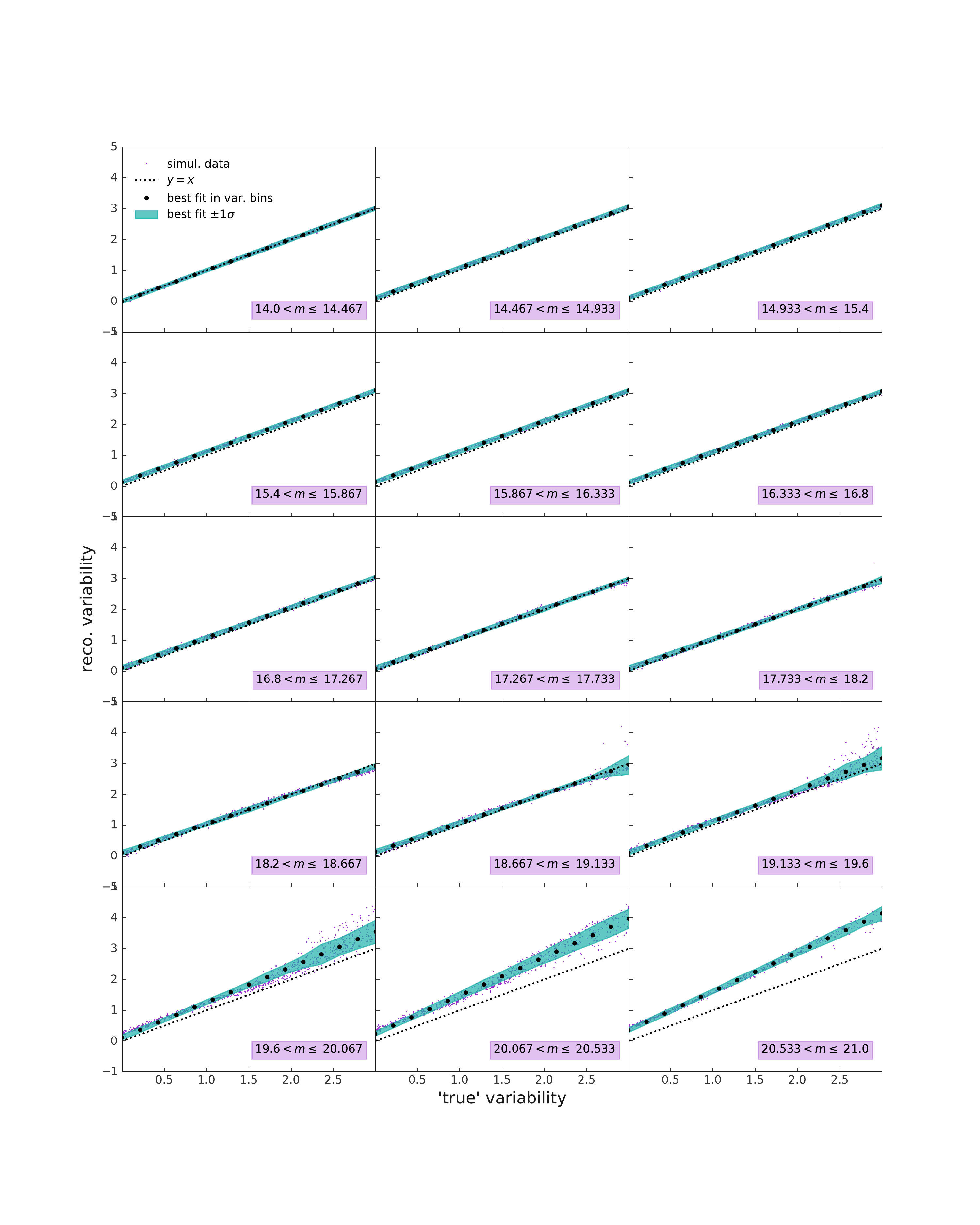}
    \caption{Example plots of the relation between the true variability and the reco variability, in various magnitude bins. The purple points are the simulated datapoints; the black points are the best fit line through the datapoints; the teal shaded region shows the 1$\sigma$ area around the best fit; the black dashed line shows the $y=x$ line for illustrative purposes.}
    \label{fig:bvt}
\end{figure*}

In practice, we take the observed variabilities (equivalent to reco variabilities) and match them up to the true variabilities, at a given magnitude. This is illustrated in Figure \ref{fig:bvt}, where a small number of bins is shown for clarity purposes. The actual number of bins used is larger, and was chosen to minimize scatter while also avoiding overbinning. Note that although the scatter in dimmer magnitude bins appears significant in the Figure, the actual scatter is much smaller and negligible. These true variabilities are injected back in to analysis.


\bsp	
\label{lastpage}
\end{document}